\documentclass[a4paper,11pt]{article}
\pdfoutput=1 

\usepackage{jcappub} 

\usepackage[T1]{fontenc} 

\usepackage{hyperref}
\usepackage[dvipsnames]{xcolor}
\usepackage{amsmath}
\usepackage{amssymb}
\usepackage{mathtools}
\usepackage{siunitx}

\usepackage{feynmp-auto}

\usepackage{cleveref}

\usepackage{float}
\usepackage{subcaption}
\newcommand{\LFV}{\Lambda_{\rm FV}}

\DeclareSIUnit{\parsec}{\text{pc}}

\title{Probing flavor violation and baryogenesis via primordial gravitational waves}

\author[a]{Zafri A. Borboruah,}
\author[b]{Anish Ghoshal,}
\author[c]{Seyda Ipek}

\affiliation[a]{Department of Physics, IIT-Bombay, Powai, Mumbai, Maharashtra, India}
\affiliation[b]{Institute of Theoretical Physics, Faculty of Physics, University of Warsaw, \\ ul. Pasteura 5, 02-093 Warsaw, Poland}
\affiliation[c]{Department of Physics, Carleton University, 1125 Colonel By Drive, Ottawa, ON, Canada}

\emailAdd{zafri123@iitb.ac.in}
\emailAdd{anish.ghoshal@fuw.edu.pl}
\emailAdd{Seyda.Ipek@carleton.ca}

\abstract{We show that observations of primordial gravitational waves of inflationary origin can shed light into the scale of flavor violation in a flavon model which also explains the mass hierarchy of fermions. The energy density stored in oscillations of the flavon field around the minimum of its potential redshifts as matter and is expected to dominate over radiation in the early universe. At the same time, the evolution of primordial gravitational waves acts as bookkeeping to understand the expansion history of the universe. Importantly, the gravitational wave spectrum is different if there is an early flavon dominated era compared to radiation domination expected from a standard cosmological model and this spectrum gets damped by the entropy released in flavon decays, determined by the mass of the flavon field $m_S$ and new scale of flavor violation $\Lambda_{\rm FV}$. We derive analytical expressions of the frequency above which the spectrum is damped, as-well-as the amount of damping, in terms of $m_S$ and $\LFV$. We show that the damping of the gravitational wave spectrum  would be detectable at BBO, DECIGO, U-DECIGO, $\mu-$ARES, LISA, CE and ET detectors for $\Lambda_{\rm FV}=10^{5-10}$ GeV and $m_S=\mathcal{O({\rm TeV})}$. Furthermore, the flavon decays can source the baryon asymmetry of the universe. We identify the $m_S-\LFV$ parameter space where the observed baryon asymmetry $\eta \sim 10^{-10}$ is produced and can be tested by gravitational wave detectors like LISA and ET. We also discuss our results in the context of the recently measured stochastic gravitational background signals by NANOGrav. 

}

\makeatletter
\gdef\@fpheader{}
\makeatother

\begin{document}
\maketitle
\flushbottom

\section{Introduction}

Our observations about the earliest moments of the universe only go back to the era of Big Bang Nucleosynthesis (BBN), at temperatures $T\sim $ a few MeV.  It is very challenging if not impossible to test pre-BBN universe via traditional cosmological observations like CMB, LSS, 21-cm etc. One very exciting novel avenue to probe the cosmological history of our universe up to much earlier times is through primordial gravitational wave (GW) observations. 

One of the most popular cosmological models for the origin of our universe involves a period of inflation. Inflation solves the horizon and flatness problems and generates primordial density fluctuations seeding the observed large scale structure. (See Ref. \cite{Martin:2013tda} for a comprehensive review.) Inflation also generates stochastic primordial GWs \cite{Grishchuk:1974ny,Starobinsky:1979ty,Rubakov:1982df} (see also \cite{Guzzetti:2016mkm} for a review). Observation of this primordial GW spectra can shed light on the details of inflationary models through, for example, their dependence on the tensor-to-scalar ratio $r$ and the reheating temperature after inflation~\cite{Bernal:2020ywq,Nakayama:2008ip,Nakayama:2008wy,Kuroyanagi:2011fy,Buchmuller:2013lra,Buchmuller:2013dja,Jinno:2014qka,Kuroyanagi:2014qza}. Importantly, the time evolution of the Hubble rate affects how GWs at different frequencies are red-shifted to the present day. As such, primordial GWs also act as record keepers of the expansion history of our universe~\cite{Seto:2003kc,Boyle:2005se,Boyle:2007zx,Kuroyanagi:2008ye,Nakayama:2009ce,Kuroyanagi:2013ns,Jinno:2013xqa,Saikawa:2018rcs,Barman:2023ktz,Ghoshal:2022ruy}. Of particular interest for this study is an era of early matter domination and its effects on the primordial GW spectra.

In the standard cosmological history assuming only the Standard Model (SM) particles and interactions, it is predicted that the universe was radiation dominated down to a temperature of $T\sim$ eV. On the other hand, we know that there must be beyond-the-SM (BSM) physics in order to explain, among other phenomena, the observed baryon asymmetry of the universe (BAU). The BAU is often quantified as the baryon-to-entropy ratio and is measured both from the CMB~\cite{Planck:2018jri} and the BBN~\cite{Fields:2019pfx}:
\begin{equation}
\eta_{\rm obs} = \frac{n_{B}-n_{\bar{B}}}{s} \simeq 8 \times 10^{-11}\,.
\label{etaBobs}
\end{equation}

In order to explain the BAU, a particle physics model must satisfy the three so-called Sakharov conditions
\cite{Sakharov:1967dj}:
\textbf{(i)} Baryon (or lepton) number must be violated, \textbf{(ii)} $C$ and $CP$
must be violated and \textbf{(iii)} there must be departure from thermal
equilibrium. In the SM, baryon number, $C$ and $CP$ are violated through electroweak (EW) interactions. However, the $CP$ violation in the quark mixing matrix is not nearly enough to produce the observed BAU~\cite{Gavela:1993ts}. Furthermore, there is no out-of-equilibrium process in the SM that could generate the BAU. Hence, new physics with extra sources of $CP$ violation as well as an out-of-equilibrium process is needed to explain the matter--antimatter asymmetry. 

In order to avoid dilution by exponential expansion and match observations, the BAU must have been produced sometime after inflation ends and before BBN. The new physics involved in generating this asymmetry could alter the early universe and leave observable signatures. For example, in a very popular class of models called EW \emph{baryogenesis}, the out-of-equilibrium condition is satisfied via a first-order cosmological phase transition, producing GWs detectable at future observatories like LISA~\cite{Seoane:2013qna}. Another common BSM ingredient to generate the BAU is a long-lived particle that decays out of equilibrium. The most important examples of such a scenario are \emph{leptogenesis} models which include right-handed neutrinos that also explain the neutrino masses. Long-lived particles can start dominating the energy density of the universe before they decay. This matter-dominated era leaves interesting cosmological signatures on the primordial GW spectra that can be probed~\cite{Berbig:2023yyy,Vagnozzi:2023lwo,Yu:2023lmo,Ghosh:2023tyz}.  

In this work we investigate the discovery potential of a \emph{flavon} leptogenesis model~\cite{Chen:2019wnk} through its effect on primordial GWs using current and future GW detectors. Flavon is a scalar field whose vacuum expectation value (VEV) determines the (Yukawa) couplings of the SM fermions. (The most famous example is the Froggatt-Nielsen field~\cite{Froggatt:1978nt}.) The oscillations of the flavon around the minimum of its potential redshift as matter. Hence, the energy density stored in these oscillations could start dominating over the radiation energy density in the early universe, before the flavon decays. In \cite{Chen:2019wnk} it was shown that this flavon-dominated era assists in generating the BAU by modifying the Hubble expansion rate. During such an era right-handed electrons do not equilibrate with the rest of the SM plasma down to $T\sim O(10~{\rm GeV})$. Hence, a lepton asymmetry, created by flavon decays to leptons, can be stored in right-handed electrons and turned into the baryon asymmetry via electroweak (EW) sphalerons. 

The era of flavon domination, at a temperature range $T\sim 10^8-10$~GeV, alters the primordial GW spectrum at frequencies that depend on the flavon mass $m_S$ and the effective scale of flavor violation $\LFV$, leaving its imprints on the expected GW signals at observatories such as DECIGO, SKA, a-LIGO and others\footnote{Breaking of the flavor symmetries can also lead to domain walls and their consequent annihilation can lead to GW signals~\cite{Gelmini:2020bqg}. }. (In Figure~\ref{fig:spectrum} we include a comprehensive list of detectors.) By comparing the detector sensitivities and the expected GW signal, we show that future observatories can probe a flavon mass  $m_S=O({\rm GeV - TeV})$ as well as a flavor violation scale $\LFV=O(10^{5-10}~{\rm GeV})$. Our analysis also shows that the parameter space that can successfully generate the BAU will also be probed by future experiments like LISA and ET. The spectral shape of gravitational waves within our model can be fitted with recent NANOGrav data~\cite{NANOGrav:2023gor,NANOGrav:2023hvm}, however successful baryogenesis is not possible. 

The paper is organized as follows. In Sec.~\ref{sec:CS_GWs} we discuss the baryogenesis mechanism and the flavon decay process. In Sec.~\ref{sec:distortion}, we summarize inflationary gravitational waves in presence of an intermediate flavon dominated phase and then show the signature of flavon decay in inflationary gravitational waves and future detectibility in experiments like LISA, BBO, DECIGO, U-DECIGO etc. Finally we conclude our discussion in Sec.~\ref{sec:conclusion}.

\medskip

\section{Flavon Cosmology and Baryogenesis}
\label{sec:CS_GWs}

In this section we summarize the baryogenesis mechanism introduced in \cite{Chen:2019wnk}, which will be used as the basis of our current analysis. In this mechanism, a right-handed electron asymmetry is generated through out-of-equilibrium decays of the flavon and subsequently this asymmetry is turned into the BAU by the EW sphalerons. Importantly for the current work, the reason the right-handed electron asymmetry is not washed out long before the EW sphalerons shut off is that the flavon energy density, falling off like non-relativistic matter, dominates the energy density of the universe over radiation.

We start with the specific flavon model relevant for the asymmetry generation. Consider the following simplified Froggatt-Nielsen \cite{Froggatt:1978nt} flavon $S$ which is a SM singlet charged under a flavor symmetry $U(1)_{\rm FN}$,
\begin{align}
    \mathcal{L}\supset \left(\frac{v_S+S}{\Lambda_{\rm FV}}\right)^{n_i} \overline{e}_R^i\phi^\ast \ell_L^i + {\rm h.c.}\,, \label{eq:mainLag}
\end{align}
where $\phi$ is the Higgs field, $v_S$ is the flavon vacuum expectation value (VEV) and $\Lambda_{\rm FV}$ is the cutoff scale for the flavor symmetry. (We are only considering the lepton sector.) The exponents $n_i$ are related to the flavor charge under $U(1)_{\rm FN}$. We use the benchmark values 
\begin{align}
    n_e=9\,,~~n_\tau = 3\,, ~~ \epsilon=\frac{v_S}{\Lambda_{\rm FV}}=0.2\,.
\end{align}
These interactions will lead to the following flavon decays:
\begin{align}
    S\to \bar{\ell}_L+e_R + \phi\,,~~S^\ast \to \bar{e}_R+\ell_L+\phi^\ast\,.\label{eq:flavondecays}
\end{align}
Here $\ell$ is the lepton doublet and $e_R$ is the right-handed electron singlet field.
It can easily be seen that a possible initial flavon asymmetry will result in an asymmetry between left-handed antileptons and left-handed leptons. The total lepton number $L$ is still conserved, because this asymmetry is balanced by an equal and opposite asymmetry in right-handed leptons. Even though there is no total lepton asymmetry, sphalerons only act on left-handed antileptons and so, they partially convert this left-handed lepton asymmetry into a baryon asymmetry. 

In standard cosmology, any asymmetry in SM particles would be washed out as they come into thermal equilibrium in the early universe. Right-handed electrons are the last SM particles to come into equilibrium, at a temperature $T\sim 10^5$~GeV in a radiation dominated universe. This equilibriation happens mainly through their interactions with the Higgs boson and $2\to2$ scatterings, with a rate $\Gamma_{LR}\simeq10^{-2}y_e^2 T$ \cite{Bodeker:2019ajh}, where $y_e\simeq 2.9\times 10^{-6}$ is the electron Yukawa. However, the flavon alters the evolution of the early universe and this story changes.

Flavon is a weakly coupled scalar field and as such it can have coherent oscillations around its $T=0$ expectation value. (Thermal corrections to its potential shift the flavon from its $T=0$ minimum~\cite{Lillard:2018zts}.) Assuming the universe is radiation dominated at the time, these oscillations start at a temperature $T_{\rm osc} \simeq \sqrt{m_S M_{\rm Pl}^\ast}$, where $m_S$ is the flavon mass, $M_{\rm Pl}^\ast=M_{\rm Pl}/\sqrt{2.75\, g_\ast}$ is the reduced Planck mass and $g_\ast$ is the effective degrees of freedom. For flavon masses $m_S=1-100$~TeV, $T_{\rm osc}\simeq 10^{10-11}$~GeV. Later we will take $T_{\rm RH}\simeq 10^{11}$~GeV. Here we note that these temperatures imply energy scales that are larger than $\LFV \simeq 10^{6-9}$~GeV that will be relevant for the baryon asymmetry. This means the effective Lagrangian in \Cref{eq:mainLag} would not be valid and one requires to use a UV complete model. The oscillation dynamics and the flavon yield needs to be recalculated in a UV complete model. We leave this model building exercise to future work and here assume that the flavon oscillations will proceed. We emphasize that the relevant energy scales for baryogenesis and the GW spectra are much lower than $\LFV$.

The energy density stored in flavon oscillations drops like non-relativistic matter, $\rho_S \propto a^{-3}$, where $a$ is the scale factor. Since radiation energy density falls of like $a^{-4}$, it is expected that the flavon energy density will take over, at a time $t_\ast$ corresponding to a temperature $T_\ast \leq T_{\rm osc}$, and start dominating the energy density of the universe. Later, as the flavon decays, universe will again become radiation-dominated at a temperature $T_{\rm rad}$. Assuming the decay products of flavon instantly thermalize, the evolution of the flavon and radiation energy densities, $\rho_S$ and $\rho_R$ respectively, is given by the following Boltzmann equations,
\begin{align}
\begin{split}
        \frac{d\rho_S}{dt} + 3H \rho_S&=-\Gamma_S \rho_S\,, \\
     \frac{d\rho_R}{dt} + 4H \rho_R &=\Gamma_S \rho_S\,, \label{eq:rhoSrad}
\end{split}
\end{align}
where 
\begin{align}
    \label{eq:gamS}
    \Gamma_S &\simeq \frac{1}{\epsilon}\frac{(n_\tau y_\tau)^2}{64\pi^3}\frac{m_S^3}{\Lambda_{\rm FV}^2}\simeq 2.3\times 10^{-17}\,{\rm GeV}\left(\frac{m_S}{{\rm TeV}}\right)^3\left(\frac{10^{10}~ {\rm GeV}}{\LFV}\right)^2\,,\notag\\ 
    H^2&=\frac{8\pi}{3M_{\rm Pl}^2}(\rho_S+\rho_R)\,,
\end{align} 
are the flavon decay width and the Hubble parameter respectively and $y_\tau\simeq 0.01$ is the tau Yukawa. The two Boltzmann equations can be solved numerically with the initial conditions $\rho_S(t_\ast)=\rho_R(t_\ast)$.

\begin{figure}[t]
\centering
\includegraphics[width=.49\linewidth]{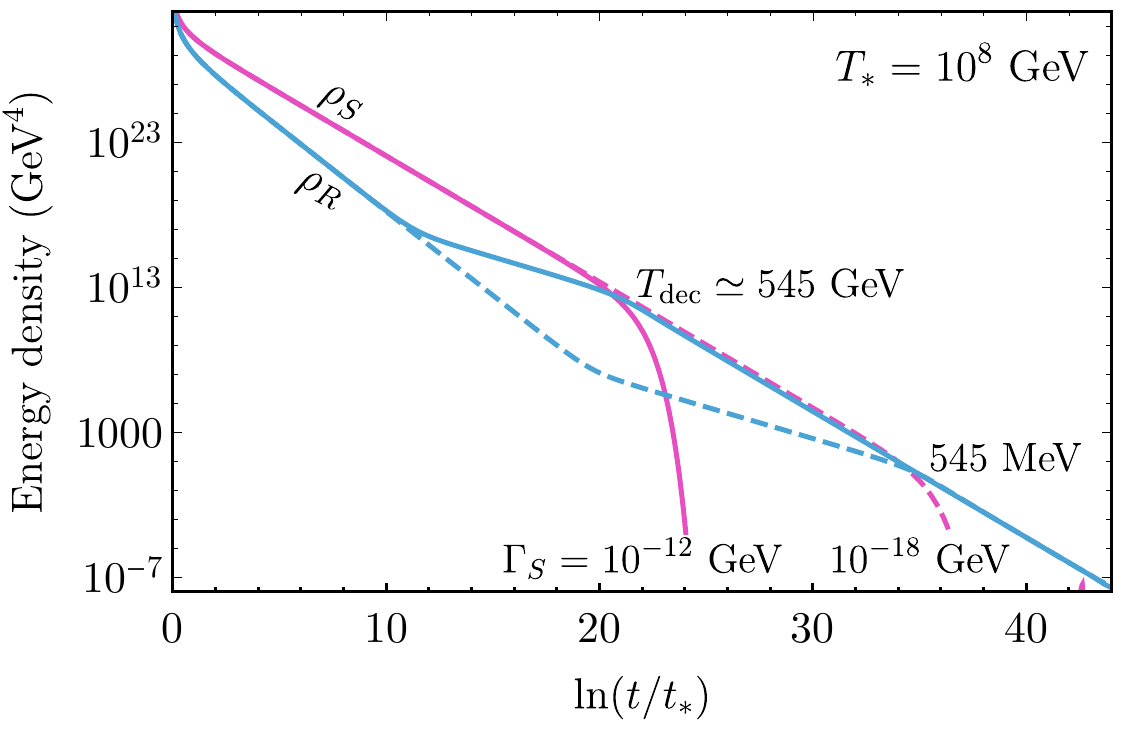}\hfill
\includegraphics[width=.49\linewidth]{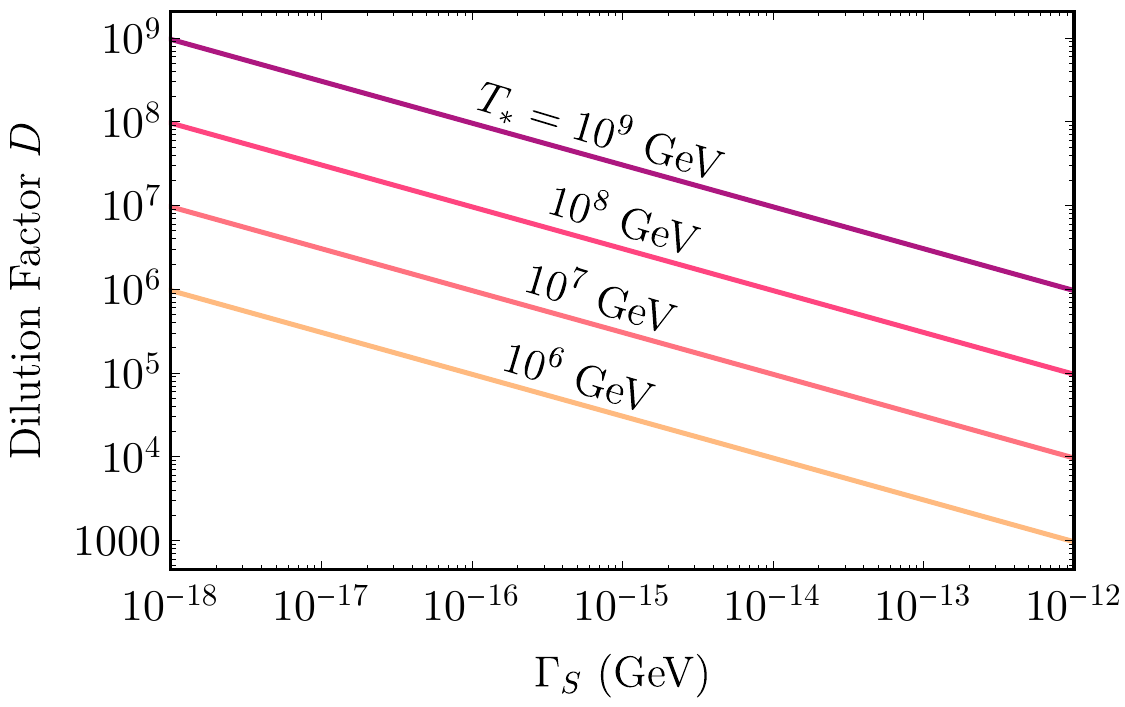}
\caption{\small{\textbf{(Left)} Evolution of flavon ($\rho_S$) and radiation ($\rho_R$) energy densities, assuming that the flavon energy density starts dominating the energy density of the universe at time $t_\ast$, corresponding to a temperature $T_\ast=10^8$~GeV, for flavon decay width $\Gamma_S=10^{-12}$~GeV (solid) and $10^{-18}$~GeV (dashed).  We also note the temperature $T_{\rm dec.}$ when flavon decays and the universe goes back to radiation domination. \textbf{(Right)} The dilution factor as defined in Eq. \eqref{eq:dilution_factor}.}}\label{fig:rhoT}
\end{figure}

In Figure \ref{fig:rhoT} we show the evolution of these energy densities for a range of parameters that will be relevant for both baryogenesis and GW observations. The temperature at which radiation starts dominating again, $T_{\rm dec}$, can be estimated by assuming sudden decay of the flavon, $\Gamma_S =H(T_{\rm dec})$, and is given by
\begin{align}
    T_{\rm dec}\simeq 1.8~{\rm GeV}\sqrt{\frac{\Gamma_S}{10^{-17}~{\rm GeV}}}\simeq 2.7~{\rm GeV}\left(\frac{m_S}{{\rm TeV}}\right)^{3/2}\left(\frac{10^{10}~{\rm GeV}}{\LFV}\right)\,, \label{eq:Tdec}
\end{align}
for $g_\ast=106.75$. Note that this temperature is independent of $T_\ast$, the temperature when flavon energy density takes over.

The decay of flavon particles into radiation increases the total entropy of the Universe. This entropy dump along with the universe entering radiation domination from matter-domination driven by flavon decays determines the wavenumber at which the primordial GW spectrum is suppressed. The dimensionless dilution factor, $D$, that will be used in the next section is the ratio between the entropy per co-moving volume just before and just after the entropy injection process. It is given by~\cite{Scherrer:1984fd}
\begin{align}
\label{eq:dilution_factor}
    D &= \frac{s(T_\text{after})a^3(T_\text{after})}{s(T_\text{before})a^3(T_\text{before})}=\left(1+2.95\left(\frac{2\pi^2 \left<g_*(T)\right>}{45}\right)^{1/3}\frac{(\frac{\rho_S}{s}|_{\rm initial})^{4/3}}{(M_\text{Pl}\,\Gamma_S)^{2/3}}\right)^{3/4} \\
    & \simeq 2\times 10^6\left(\frac{T_\ast}{10^6~{\rm GeV}}\right)\left(\frac{\LFV}{10^{10}~{\rm GeV}}\right)\left(\frac{{\rm TeV}}{m_S}\right)^{3/2}\,, \notag
\end{align}
where $s$ is the entropy density and $a$ is the scale factor. Here we assume $\left<g_*(T)\right>\sim g_*\sim 106.75$. $T_\text{before/after}$ represent temperature before/after the decay process and $\frac{\rho_S}{s}|_{\rm initial}$ is calculated at the temperature $T_*$. The numerical estimate in the second line is obtained by using the initial flavon energy density  and initial entropy density,
\begin{equation}
    \label{eq:initial_entropy}
      \rho_S|_{\rm initial} = \frac{\pi^2}{30}g_\ast T_\ast ^4\,,~~s|_{\rm initial}=\frac{2\pi^2}{45}g_{*s}T_\ast^3\,.
\end{equation}
where $g_{*s}$ is the relativistic degrees of freedom contributing to entropy~\cite{Kolb:1990vq} which we assume to be equal to $g_*\sim 106.75$ at $T_*$.
In Figure~\ref{fig:rhoT} we show the evolution of the radiation and flavon energy densities as well as the dilution factor for different model parameters.

\noindent\emph{\textbf{Baryon Asymmetry From Flavon Decays}}

\begin{figure}[t]
\centering

\includegraphics[width=.49\linewidth]{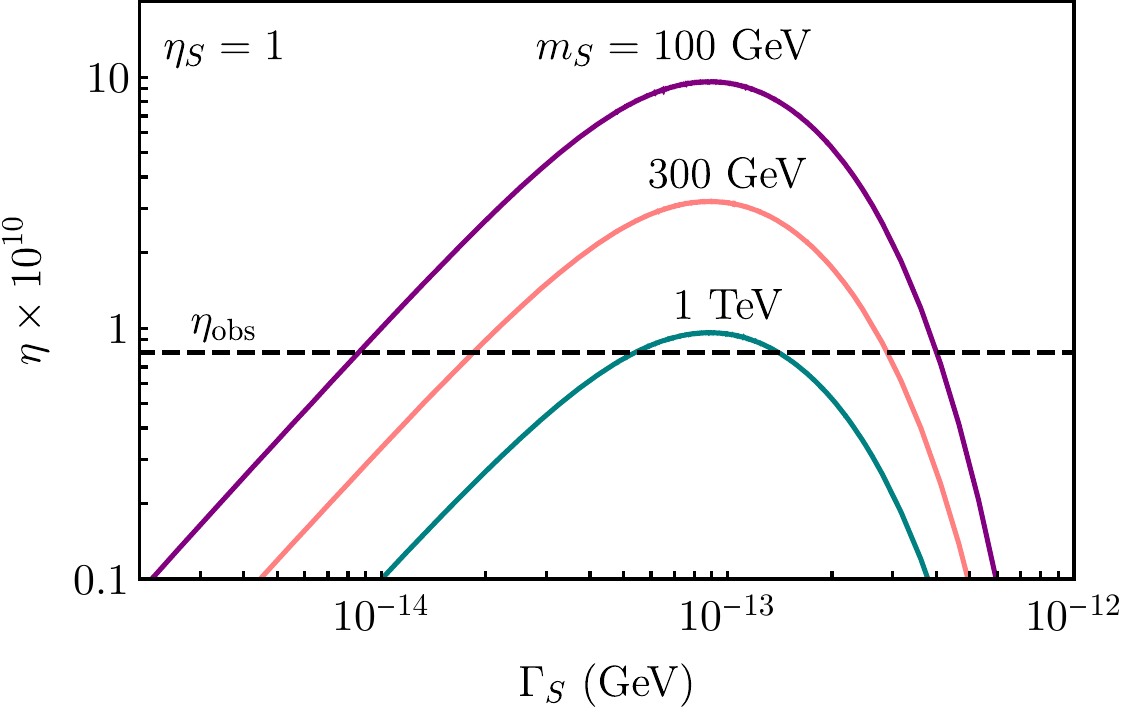}
    \includegraphics[width=.49\linewidth]{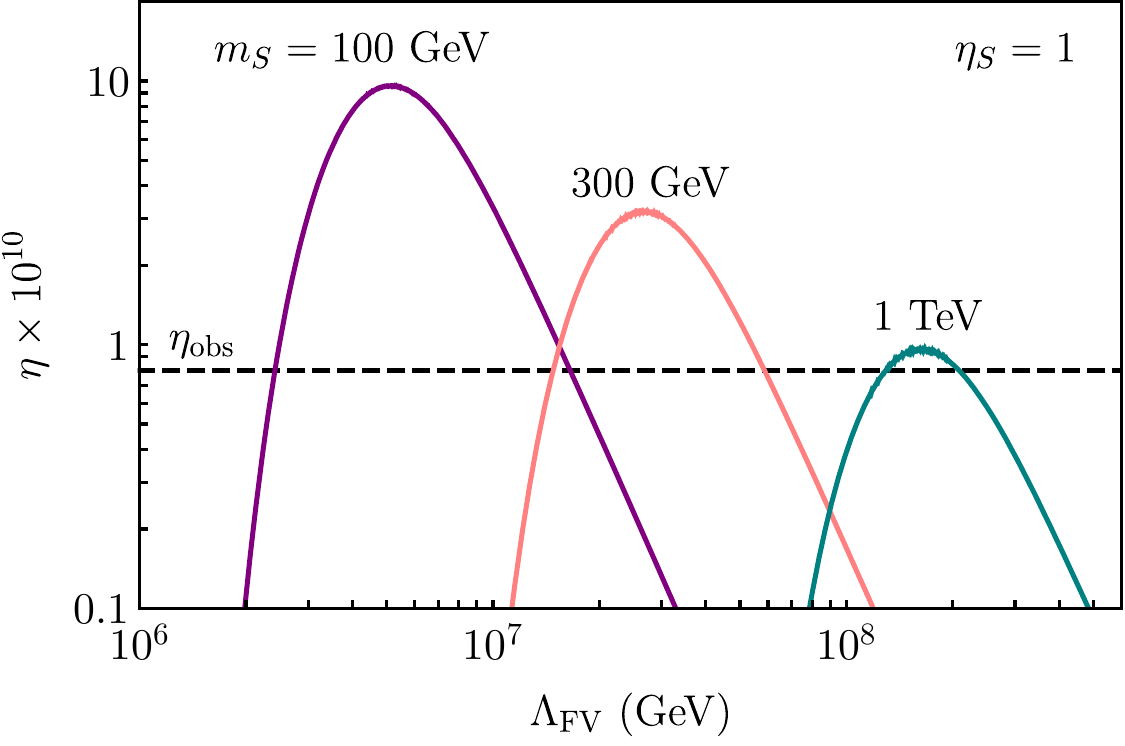}

\caption{ \small{Baryon asymmetry of the universe versus the flavon decay width $\Gamma_S$ (left) and the scale of flavor symmetry breaking $\Lambda_{\rm FV}$ (right). (Note that the baryon asymmetry is independent of the flavon-domination temperature $T_\ast$.) The horizontal dashed line is the observed baryon asymmetry, $\eta\simeq 0.8\times 10^{-10}$. We assumed an initial flavon asymmetry of $\eta_S=1$. The baryon asymmetry decreases with increasing flavon mass $m_S$ simply because the number density of flavons decreases.}}\label{fig:etaGS}
\end{figure}

As stated above, flavons decay into right-handed electrons while anti-flavons decay into right-handed anti-electrons. (See Equation~\ref{eq:flavondecays}.) Assuming an initial flavon asymmetry $\eta_S$, we can track the asymmetry in the number densities of right-handed electrons and right-handed anti-electrons via the following Boltzmann equation:
\begin{align}
    \frac{d\Delta_{e_R}}{dt} = -3 H \Delta_{e_R} -\Gamma_{LR} \Delta_{e_R} + B_e\Gamma_S \Delta_S\,, \label{eq:eRasym}
\end{align}
where $\Delta_{e_R} = n_{e_R}-n_{\bar{e}_{R}}$ is the asymmetry in right-handed electron number density, $B_e \simeq 7.5\times10^{-7}$ is the flavon branching fraction to electrons and 
\begin{align}
    \Delta_S = \eta_S \frac{\rho_S}{m_S}\,,
\end{align}
is the flavon asymmetry. Equation~\ref{eq:eRasym} is solved together with Equation~\ref{eq:rhoSrad} to find the right-handed electron asymmetry. This asymmetry is turned into a baryon asymmetry by the EW sphalerons, which are in thermal equilibrium at temperatures $T_{\rm EW}\gtrsim 160$~GeV. The final baryon asymmetry is given by \cite{Planck:2015fie,Riemer-Sorensen:2017vxj}
\begin{align}
    \eta\equiv \frac{n_B - n_{\bar{B}}}{s} \simeq \frac{198}{481}\frac{\Delta_{e_R}(T=T_{\rm EW})}{s}\,,
\end{align}
where $s$ is the entropy density of the universe. In Figure~\ref{fig:etaGS} we show the final baryon asymmetry for varying $\Gamma_S,\, m_S$ and $\LFV$. We take the initial flavon asymmetry $\eta_S=1$ but note that in parts of the parameter space $\eta_S \sim O(0.1)$ can be accommodated. Successful leptogenesis is achieved for a flavor violation scale in the range $O(10^{6-9}~{\rm GeV})$ for flavon masses $m_S \lesssim 1$~TeV. Larger flavon domination temperature, $T_\ast$, results in a larger right-handed electron asymmetry $\Delta_{e_R}$ as well as larger dilution when the flavon decays. As a result, the final baryon-to-entropy ratio $\eta$ is independent of $T_\ast$.  

\section{Primordial Gravitational Waves and flavon domination}\label{sec:distortion}

In this section we analyze the inflationary-origin GW spectrum including an era of flavon domination. We start with a review of the power spectrum expected at GW detectors originating from tensor perturbations during inflation. We then show an era of flavon domination leaves observable imprints on this spectrum.

\subsection{Distortion of the inflationary tensor mode spectrum}

Primordial GWs, seeded by tensor perturbations of the metric, are predicted in all inflationary models. These GWs generated during inflation are scale invariant and undergo damping upon horizon re-entry. Their contemporary power spectrum can be expressed as a function of the wavenumber $k=2\pi f$:
\begin{align}
    \Omega_{\rm GW}(k)=\frac{1}{12}\left(\frac{k}{a_0 H_0}\right)^2 T^2_T(k) P_T(k)\,, \label{eq:omegaGW}
\end{align}
 where  $a_0=1$ and $H_0\simeq \qty{2.2e-4}{\per\mega\parsec}$ are the current scale factor and the Hubble expansion rate, respectively. We will use the power-law parametrization for the tensor power spectrum~\cite{Barman:2023ktz}
\begin{align}
    P_T(k)= r A_s(k_\ast)\left(\frac{k}{k_\ast}\right)^{n_T}\,,
\end{align}
where $r<0.035$~\cite{BICEP:2021xfz} is the tensor-to-scalar ratio and $A_s(k_\ast)=2.0989\times 10^{-9}$ \cite{Planck:2018jri} is the amplitude of scalar perturbations given at the pivot scale $k_\ast= \SI{0.05}{\per\mega\parsec}$. We will treat the tensor spectral index $n_T$ as a free parameter. In the standard single-field slow-roll inflation, a red-tilted spectra, $n_T<0$, is expected, according to the consistency relation $n_T = - r /8$ $\approx$ \cite{Liddle:1993fq}. However, blue-tilted spectra are plausible in various models, including string gas cosmology, super-inflation models, G-inflation, non-commutative inflation, and scenarios involving particle production during inflation among several other scenarios \cite{Brandenberger:2006xi,Baldi:2005gk,Kobayashi:2010cm,Calcagni:2004as,Calcagni:2013lya,Cook:2011hg,Mukohyama:2014gba,Kuroyanagi:2020sfw}.

In Equation \ref{eq:omegaGW},  $T_T^2(k)$ serves as a transfer function describing the evolution of gravitational waves in the background of a Friedmann-Lemaitre-Robertson-Walker universe. It is given by~\cite{Turner:1993vb,Chongchitnan:2006pe,Nakayama:2008wy,Nakayama:2009ce,Kuroyanagi:2011fy,Kuroyanagi:2014nba}:
\begin{align}
    T_T^2(k) &= \Omega_m^2 \left(\frac{g_*(T_\text{in})}{g_*^0}\right)\left(\frac{g_{*s}^0}{g_{*s}(T_\text{in})}\right)^{4/3} \left(\frac{3j_1(k\tau_0)}{z_k}\right)^2 F(k),
\end{align}
where $\Omega_m =0.31$ is the total matter density, $j_1(z_k)$ is the first spherical Bessel function, and $\tau_0=2 / H_0$ is the conformal time today. Here $g_*(T)$ and $g_{*s}(T)$ are the effective massless degrees of freedom contributing to energy and entropy densities respectively at temperature $T$~\cite{Kuroyanagi:2020sfw,Kolb:1990vq} with current values $g_*^0=3.36$ and $g_{*s}^0=3.91$. $T_{\rm in}$ is the temperature at the time of horizon re-entry \cite{Nakayama:2008wy},
\begin{equation}
   T_\text{in} = \SI{5.8e6}{\giga\electronvolt} \left(\frac{106.75}{g_*(T_\text{in})}\right)^{1/6} \left(\frac{k}{\SI{e4}{\per\mega\parsec}}\right).
\end{equation}

The factor $F(k)$ incorporates the expansion of the universe and is affected by the presence of intermediate matter domination (IMD). Without IMD, it is given by:
\begin{align}\label{eq:stand}
    F(k)_\text{standard}  =   T_1^2\left(\frac{k}{k_\text{eq}}\right)T_2^2\left(\frac{k}{k_\text{RH}}\right),
\end{align}
where 
\begin{align}
    \begin{split}
        k_\text{RH} &= \SI{1.7e14}{\per\mega\parsec} \left(\frac{g_{*s}(T_\text{RH})}{g_{*s}^0}\right)^{1/6} \left(\frac{T_\text{RH}}{10^7\;\text{GeV}}\right)\,, \\
         k_\text{eq} &= \SI{7.1e-2}{\per\mega\parsec}\, \Omega_m h^2\,, 
    \end{split}
\end{align}
correspond to the wavenumbers entering the horizon at the reheating temperature $T_{\rm RH}$ and during the standard matter domination, respectively, and $h=0.7$.

For an era of intermediate matter domination, the factor $F(k)$ is instead given by
\begin{align}
    F(k)_\text{IMD} &=  
      T_1^2\left(\frac{k}{k_\text{eq}}\right)T_2^2\left(\frac{k}{k_\text{S}}\right)T_3^2\left(\frac{k}{\tilde{k}_\text{S}}\right)T_2^2\left(\frac{k}{k_\text{RH, S}}\right)\,, \label{eq:FIMD}
\end{align}
where $k_{\rm RH,S} = k_{\rm RH}D^{-1/3}$ and 
\begin{align}
     k_\text{S} &= \SI{1.7e14}{\per\mega\parsec} \left(\frac{g_{*s}(T_\text{dec})}{g_{*s}^0}\right)^{1/6} \left(\frac{T_\text{dec}}{10^7\;\text{GeV}}\right)\,,\quad \quad \tilde{k}_\text{S}= k_\text{S}\, D^{2/3}\,.
\end{align}
These correspond to the relevant wavenumbers at the time of flavon decay and flavon domination respectively. The dilution factor $D$ accounts for the entropy dump due the flavon decays. It is given in Equation \ref{eq:dilution_factor} and shown in Figure \ref{fig:etaGS} for various model parameters.

In Equation~\ref{eq:FIMD} we use the fitting functions $T_i^2(x)$ given in \cite{Kuroyanagi:2014nba,Berbig:2023yyy}:
\begin{align}
    \begin{split}
        T_1^2(x) &= 1+1.57 x +3.42 x^2\,,\\
     T_2^2(x) &= (1-0.22x^{3/2}+0.65x^2)^{-1}\,,\\
     T_3^2(x) &= 1+0.59 x +0.65 x^2\,.
    \end{split}
\end{align}

\begin{figure}
\centering
\begin{subfigure}[b]{.85\textwidth}
   \includegraphics[width=1\linewidth]{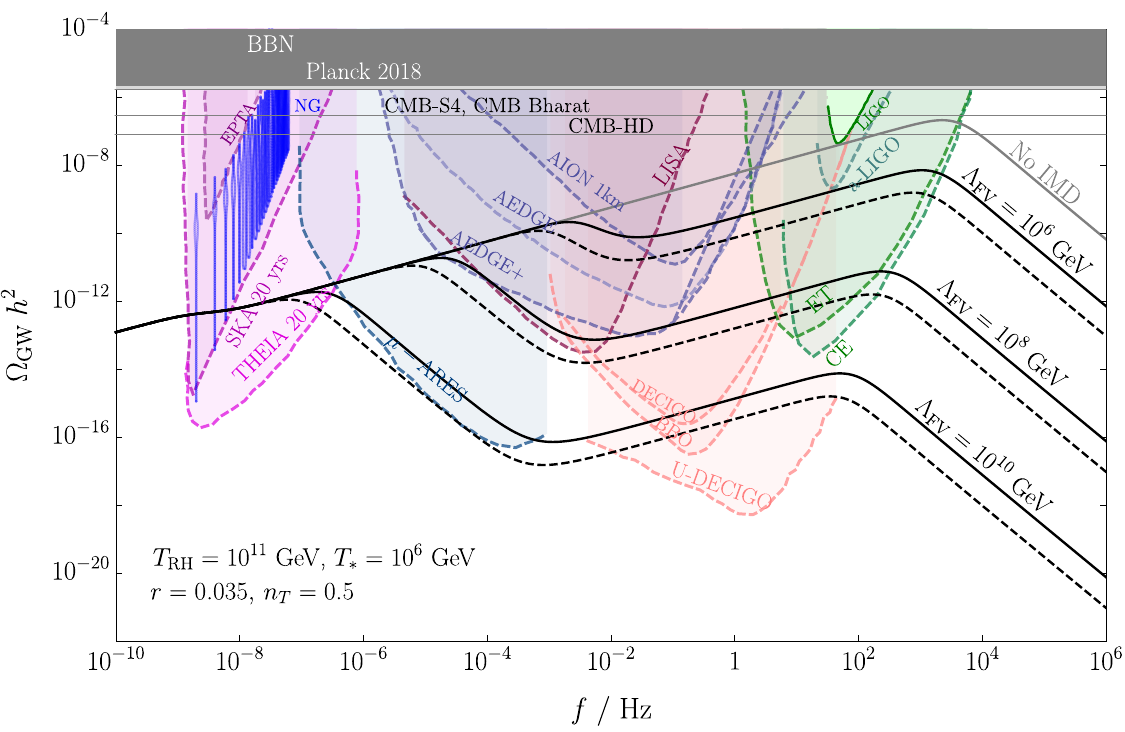}
   \label{fig:plotLam1} 
\end{subfigure}

\begin{subfigure}[b]{0.85\textwidth}
   \includegraphics[width=1\linewidth]{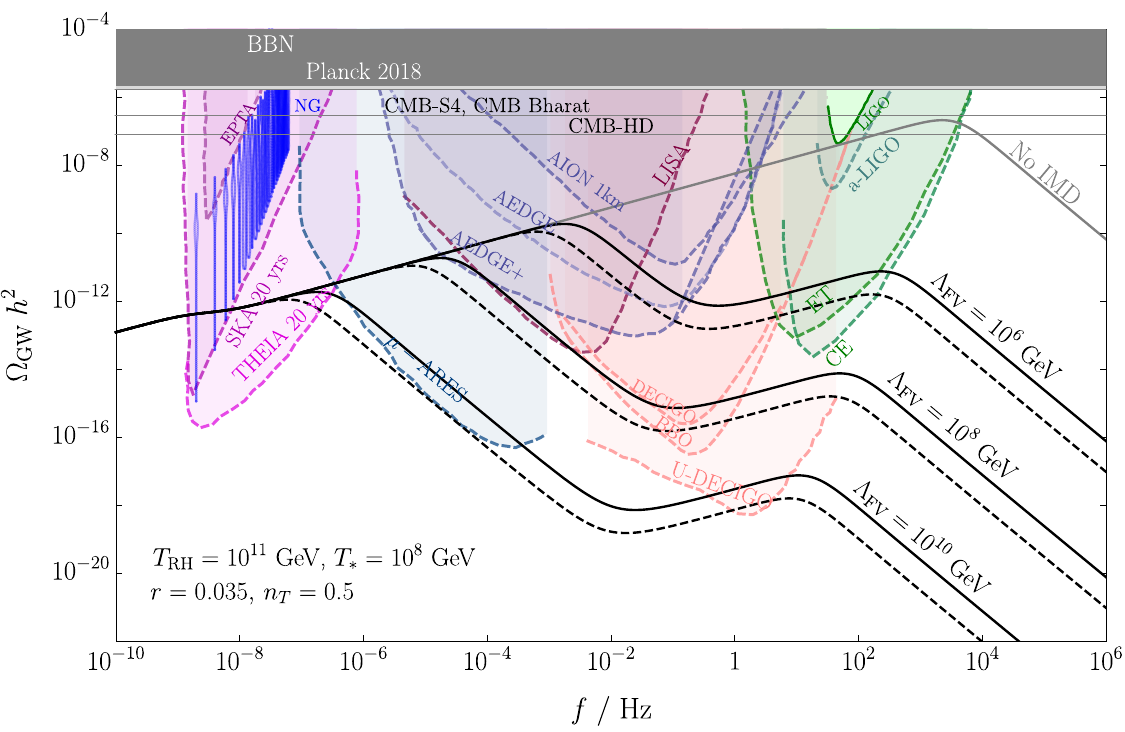}
   \label{fig:plotLam2}
\end{subfigure}
\caption{\small{Primordial GW spectra showing the suppression due to an era of flavon domination compared to the standard evolution labeled ``No IMD". We show the suppression for different values of flavor scale $\LFV=10^{6-10}$~GeV as well as for flavon masses $m_S$=2 TeV (solid) and $m_S=1$ TeV (dashed). We use two flavon domination scales, $T_*=10^6$ GeV \textbf{(top)} and $10^8$ GeV \textbf{(bottom)}. For both plots we take $T_{\rm RH}=10^{11}$~GeV, $r=0.035$ and $n_T=0.5$. The shaded regions are current and future sensitivity curves of various GW experiments. The topmost (dark) gray region is constraints from $\Delta N_{\rm eff}$. The blue violin lines are obtained from the NANOGrav 15 year dataset~\cite{NANOGrav:2023hvm,NANOGrav:2023gor}
\label{fig:spectrum}}}
\end{figure}



\begin{figure}
\centering
\includegraphics[width=0.85\linewidth]{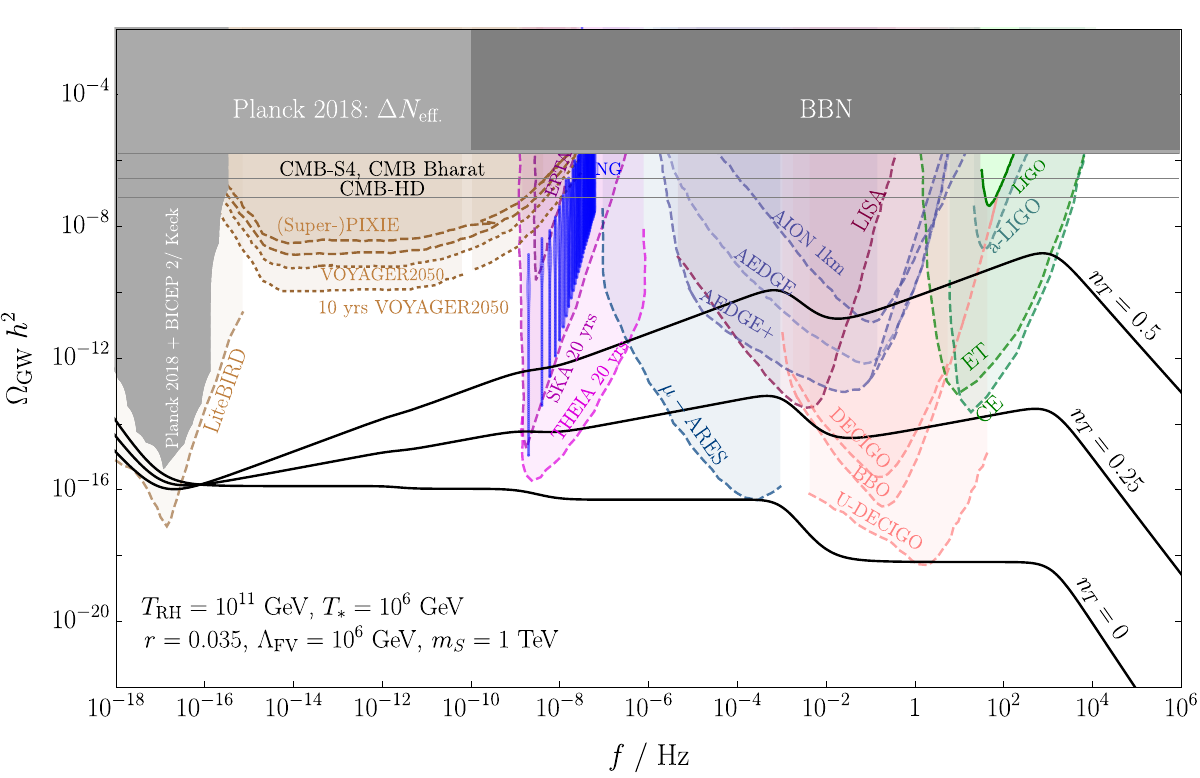}
\caption{\small{Same as Fig.~\ref{fig:spectrum} but showing the effect of $n_T$. Here we have taken $\LFV=10^6$ GeV, $m_S=1$ TeV, $T_*=10^8$ GeV and $T_{\rm RH}=10^{11}$ GeV. The future CMB experiment LiteBIRD~\cite{Hazumi:2019lys} will be able to probe different values of $n_T$ providing complementarity to other types of GW experiments.}
\label{fig:spectrumnT}}
\end{figure}

At this point, we can plot the primordial GW spectrum $\Omega_{\rm GW}h^2$ in Equation \ref{eq:omegaGW} and discuss its features accounting for an era of flavon domination. In Figure~\ref{fig:spectrum} we show the effects changing the scale of flavor violation, $\LFV$, on the power spectrum for fixed values of the flavon mass, $m_S=1$ and $2$~TeV for a blue-tilted spectrum $n_T=0.5$. It can be seen that larger $\LFV$ suppresses the GW power spectrum more. This is due to the fact that as $\LFV$ gets larger, the flavon is longer lived and hence the dilution factor is larger. Similarly, it can be seen in Figure~\ref{fig:spectrum} that lower $m_S$ suppresses the power spectrum more. This is again due to larger dilution from a longer lived flavon. Furthermore, as expected, if the flavon starts dominating earlier, again the suppression is larger. This can be seen from the difference between $T_\ast =10^8$~GeV and $T_\ast =10^6$~GeV. In Fig.~\ref{fig:spectrumnT} we show the dependence of the spectrum on $n_T$. Although a negative or small positive $n_T\sim0$ is more natural in inflationary scenarios, a large positive $n_T\sim\mathcal{O}(1)$ is not currently ruled out. Such a spectrum can also account for the recent NANOGrav signals for $n_T \sim 2.61$ as we will discuss in Sec.~\ref{sec:NG}. We can also infer from Fig.~\ref{fig:spectrum}, the spectrum for $\Lambda_{\rm FV}=10^6$ GeV, $m_S=2$ TeV, $T_*=10^6$ GeV, should be detectable at LISA, adv-LIGO, CE and ET, for instance. We illustrate the fact that due the presence of a wide range of frequency dependence with very specific features in GW spectral shapes we expect the GW signals simultaneously to be measured and reconstructed in multiple frequency bands corresponding to various GW detectors running at the same time. At this stage these microphysical BSM parameters may or may not correspond to successful baryogenesis. We will explore the baryogenesis consequences in later sections.

The suppression of the primordial gravitational wave spectrum due to flavon domination happens above a frequency $f_{\rm sup}$ determined by the time of flavon decay given in Equation~\ref{eq:Tdec}: 

\begin{align}
    f_\text{sup}
    &\simeq 2.7\times10^{-8}\;\text{Hz}\; \left(\frac{T_\text{dec}}{{\rm GeV}}\right)\simeq 7.3\times 10^{-8}\,{\rm Hz}\left(\frac{m_S}{{\rm TeV}}\right)^{3/2}\left(\frac{10^{10}~{\rm GeV}}{\LFV}\right)\,.\label{eq:freq}
\end{align}
The suppression factor of the power spectrum, denoted as $R_\text{sup}$, is given by \cite{Seto:2003kc}:
\begin{equation}
    R_\text{sup} = \frac{\Omega_\text{GW}^\text{IMD}}{\Omega_\text{GW}^\text{standard}}\simeq \frac{1}{ D^{4/3}} \simeq 8.7\times 10^{-8}\left(\frac{10^6~{\rm GeV}}{T_\ast}\right)^{4/3}\left(\frac{10^{10}~{\rm GeV}}{\LFV}\right)^{4/3} \left(\frac{m_S}{{\rm TeV}}\right)^{4/3}\,, \label{eq:Rsup}
\end{equation}
where $\Omega_\text{GW}^\text{IMD}$ is the power spectrum with an era flavon domination, while $\Omega_\text{GW}^\text{standard}$ is the one in the standard scenario without intermediate matter domination. The suppression factor depends on the flavon domination temperature $T_\ast$, flavon mass $m_S$ and the effective scale of flavor violation $\LFV$ via the dilution factor $D$ in \ref{eq:dilution_factor}.

We also display the sensitivity curves for a plethora of current and future GW detectors in Figure \ref{fig:spectrum}. These experiments can be grouped as follows.
\begin{itemize}
    \item \textbf{ground based  interferometers:} \textsc{LIGO}/\textsc{VIRGO}             \cite{LIGOScientific:2016aoc,LIGOScientific:2016sjg,LIGOScientific:2017bnn,LIGOScientific:2017vox,LIGOScientific:2017ycc,LIGOScientific:2017vwq}, a\textsc{LIGO}/a\textsc{VIRGO}  \cite{LIGOScientific:2014pky,VIRGO:2014yos,LIGOScientific:2019lzm}, \textsc{AION} \cite{Badurina:2021rgt,Graham:2016plp,Graham:2017pmn,Badurina:2019hst}, \textsc{Einstein Telescope (ET)} \cite{Punturo:2010zz,Hild:2010id}, \textsc{Cosmic Explorer (CE)}  \cite{LIGOScientific:2016wof,Reitze:2019iox},
    \item   \textbf{space based interferometers:}  \textsc{LISA} \cite{Baker:2019nia},\textsc{BBO} \cite{Crowder:2005nr,Corbin:2005ny}, 
    \textsc{DECIGO}, \textsc{U-DECIGO}\cite{Seto:2001qf,Yagi:2011wg}, \textsc{AEDGE} \cite{AEDGE:2019nxb,Badurina:2021rgt}, \textsc{$\mu$-ARES} \cite{Sesana:2019vho}
    \item \textbf{recasts of star surveys:} \textsc{GAIA}/\textsc{THEIA} \cite{Garcia-Bellido:2021zgu}, 
    \item \textbf{pulsar timing arrays (PTA):} \textsc{SKA} \cite{Carilli:2004nx,Janssen:2014dka,Weltman:2018zrl}, \textsc{EPTA} \cite{Lentati:2015qwp,Babak:2015lua}, \textsc{NANOGRAV}~\cite{McLaughlin:2013ira,NANOGRAV:2018hou,Aggarwal:2018mgp,Brazier:2019mmu,NANOGrav:2020bcs}
    \item \textbf{CMB polarization:} Planck 2018 \cite{Akrami:2018odb} and BICEP 2/ Keck \cite{BICEP2:2018kqh} computed by  \cite{Clarke:2020bil}, \textsc{LiteBIRD} \cite{Hazumi:2019lys}, 
    \item \textbf{CMB spectral distortions:} \textsc{PIXIE}, \textsc{Super-PIXIE}  \cite{Kogut:2019vqh}, \textsc{VOYAGER2050}~\cite{Chluba:2019kpb}
\end{itemize}
The blue violin lines in the figures are the most recent data from NANOGrav. 

Gravitational waves contribute to the energy density of the universe as dark radiation and thus can be constrained by BBN and CMB in terms of $\Delta N_{\rm eff}$. The gravitational radiation produced before BBN (and CMB) must satisfy~\cite{Maggiore:1999vm}

\begin{equation}
    \int_{f_{\rm min}}^{\infty}\frac{df}{f}\Omega_{\rm GW}(f)h^2 \leq 5.6\times10^{-6}\Delta N_{\rm eff}\,,
\end{equation}
where the lower limit of the integration $f_{\rm min}\simeq10^{-10}$ Hz for BBN and $\simeq10^{-18}$ Hz for the CMB. Ignoring this frequency dependence of $\Omega_{\rm GW}$ for simplicity, the above requirement gives $\Omega_{\rm GW}\leq 5.6\times10^{-6}\Delta N_{\rm eff}$ for the gravitational wave spectra under consideration.

In Figure~\ref{fig:spectrum} we show the BBN limit of $\Delta N_{\rm eff}^{\rm BBN}\simeq 0.4$ \cite{Cyburt:2015mya} in dark gray and the limit from the combined Planck and baryon acoustic oscillation (BAO) data, $\Delta N_{\rm eff}^{\rm Planck+BAO}\simeq0.28$ \cite{Planck:2018vyg}, in light gray. These bounds will get much stricter in the future. CMB-HD anticipates $\Delta N_{\rm eff}^{\rm Proj.}=0.014$ \cite{CMB-HD:2022bsz}, CMB-Bharat can probe $\Delta N_{\rm eff}^{\rm Proj.}=0.05$ \cite{CMB-bharat}, and both CMB Stage IV \cite{doi:10.1146/annurev-nucl-102014-021908} and NASA's PICO mission \cite{Alvarez:2019rhd} aim for $\Delta N_{\rm eff}^{\rm Proj.}=0.06$. We also show these projected bounds in Figure~\ref{fig:spectrum}.

\subsection{Signal-to-noise ratio and detection prospects}


\begin{figure}[t]
\centering
    \begin{subfigure}{0.49\linewidth}
\includegraphics[width=\linewidth]{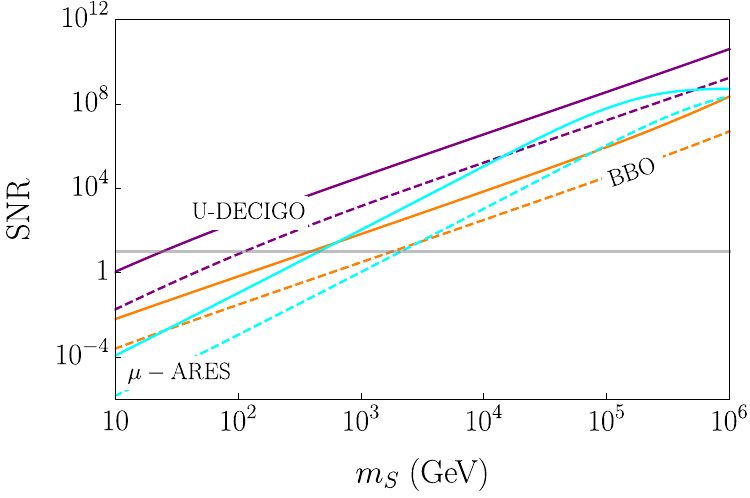} 
    \caption{\it }
    \end{subfigure}\hfill
    \begin{subfigure}{0.49\linewidth}
\includegraphics[width=\linewidth]{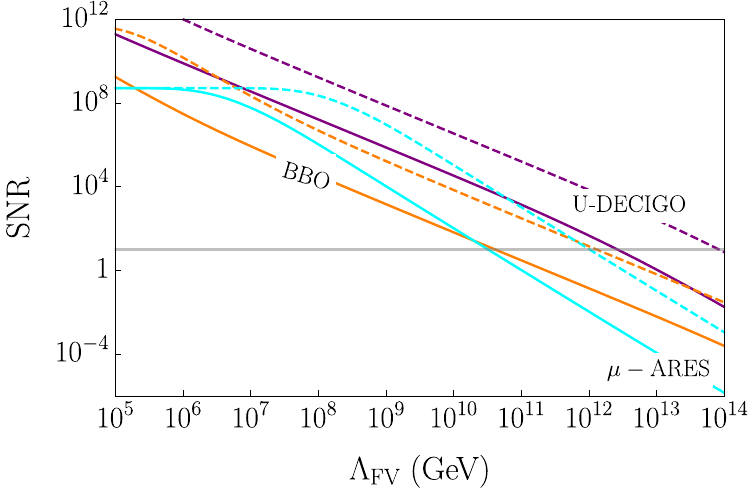}
    \caption{\it }
    \end{subfigure}\hfill
    
\caption{ \small{Variation of SNR with \textbf{(a)} $m_S$ and \textbf{(b)} $\Lambda_{\rm FV}$ at U-DECIGO (purple), BBO (orange) and ARES (cyan) experiments. In (a) the solid lines correspond to $\Lambda_{\rm FV}=10^{10}$ GeV while the dashed lines correspond to $\Lambda_{\rm FV}=10^{11}$ GeV. In (b) the solid lines correspond to $m_S=1$ TeV and dashed lines correspond to $m_S=10$ TeV. For both plots we have used $r=0.035, n_T=0.5, T_*=10^8$ GeV, and $T_{\rm RH}=10^{11}$ GeV. Gray solid lines show SNR=10, a choice of threshold for detection.}}
    \label{fig:SNRplot}
\end{figure}



\begin{figure}[ht]
\centering
    \begin{subfigure}{0.49\linewidth}
\includegraphics[width=\linewidth]{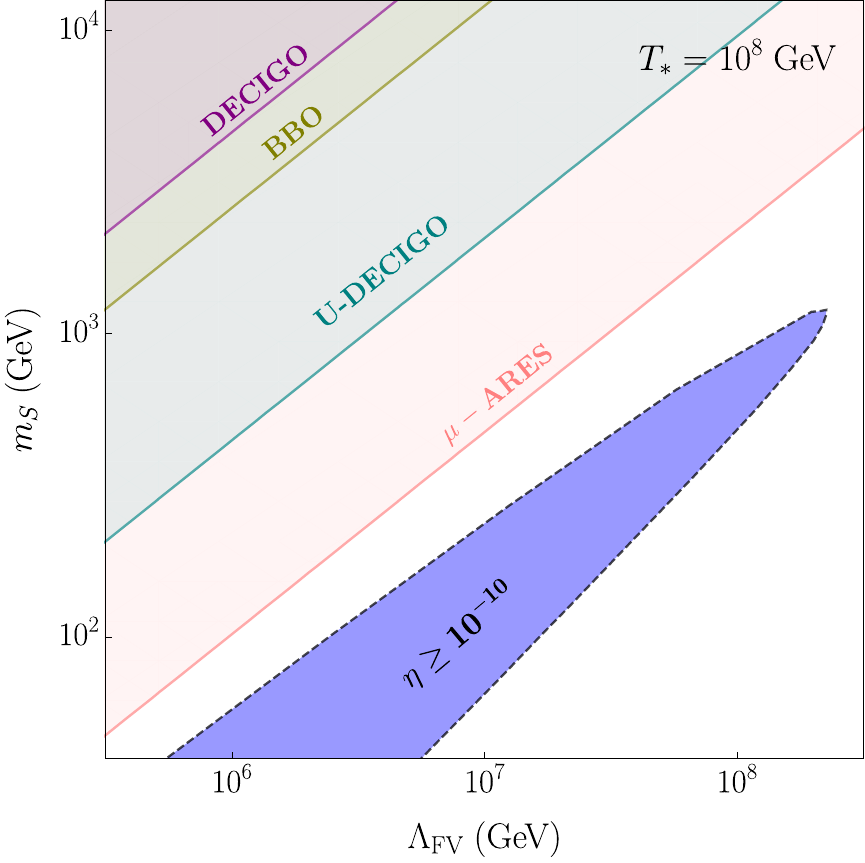} 
    \label{fig:Res1}
    \caption{\it }
    \end{subfigure}\hfill
    \begin{subfigure}{0.49\linewidth}
\includegraphics[width=\linewidth]{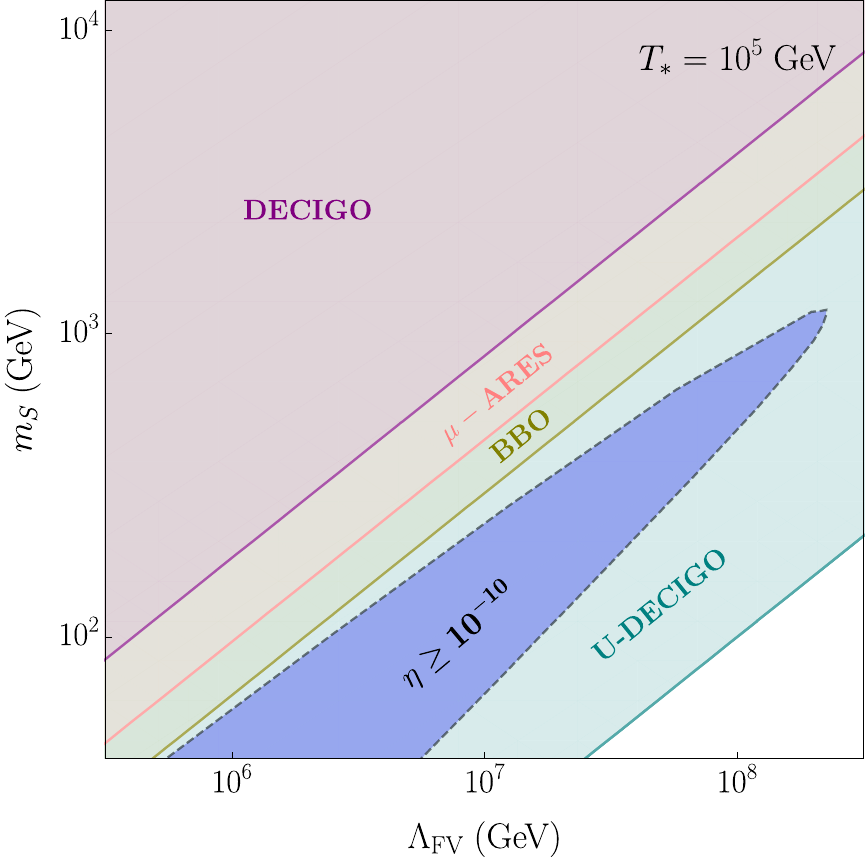}
    \label{fig:Res1Weak}
    \caption{\it }
    \end{subfigure}\hfill
    
\caption{ \small{Detection regions for various experiments, with SNR $\geq 10$, with respect to the flavon mass $m_S$ and the flavor scale $\LFV$ for \textbf{(a)} $T_*=10^8$ GeV and \textbf{(b)} $T_*=10^5$ GeV. The blue region corresponds to successful generation of baryon asymmetry, with $\eta\geq10^{-10}$.} In both plots we take $r=0.035, n_T=0$ and $T_{\rm RH}=10^{11}$~GeV.}
    \label{fig:SNR8}
\end{figure}


The sensitivity of GW experiments is given via the characteristic strain $h_\text{GW}(f)$, which can be converted into the corresponding energy density:
\begin{align}
    \Omega_\text{exp}(f) h^2 = \frac{2\pi^2 f^2}{3 H_0^2} h_\text{GW}(f)^2 h^2\,.
\end{align}
We show these energy densities for various GW experiments in Figure~\ref{fig:spectrum}. 

In order to discern the detection possibility of a GW signal, it is more useful to compute the signal-to-noise ratio (SNR) for a given experiment. We calculate SNR for a number of relevant detectors using the following definition:
\begin{align}
     \text{SNR}\equiv \sqrt{\tau \int_{f_\text{min}}^{f_\text{max}} \text{d}f \left(\frac{ \Omega_\text{GW}(f) h^2}{\Omega_\text{exp}(f) h^2}\right)^2 } \label{eq:SNR},
\end{align}
where $\tau = 4\; \text{years}$ is the observation time we have considered for each experiment and $f_{\rm min},f_{\rm max}$ are the minimum and maximum frequencies that the relevant detector is sensitive to. We set a detection threshold with $\text{SNR}\geq 10$.


\begin{figure}[ht]
\centering
   \begin{subfigure}{0.49\linewidth}
\includegraphics[width=\linewidth]{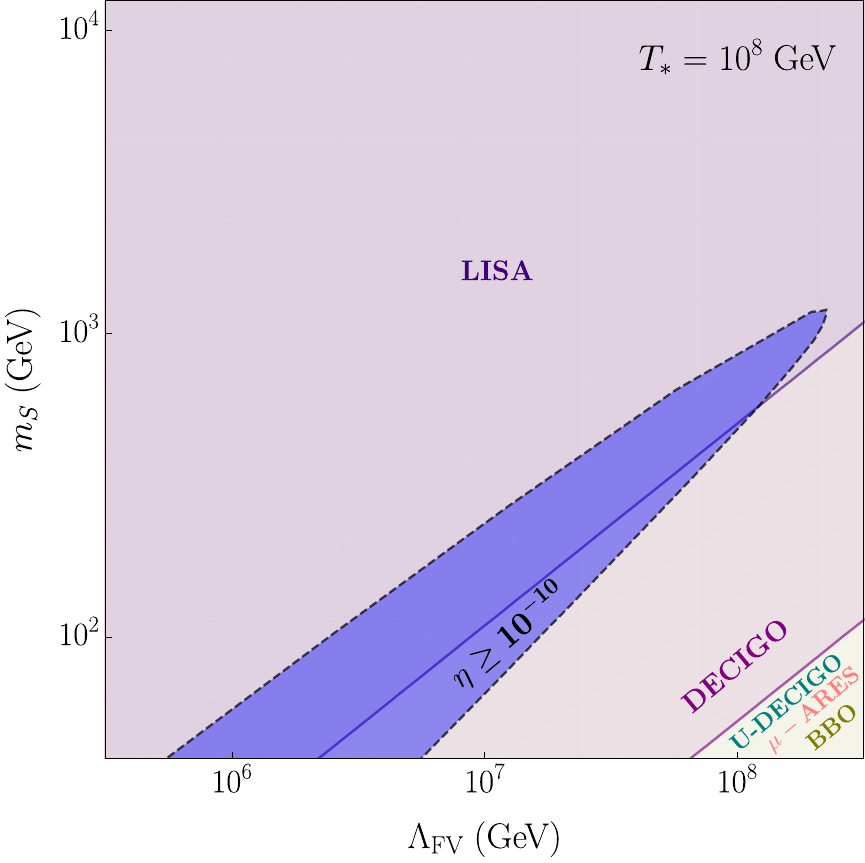} 
   \caption{\it }
   \end{subfigure}\hfill
    \begin{subfigure}{0.49\linewidth}
\includegraphics[width=\linewidth]{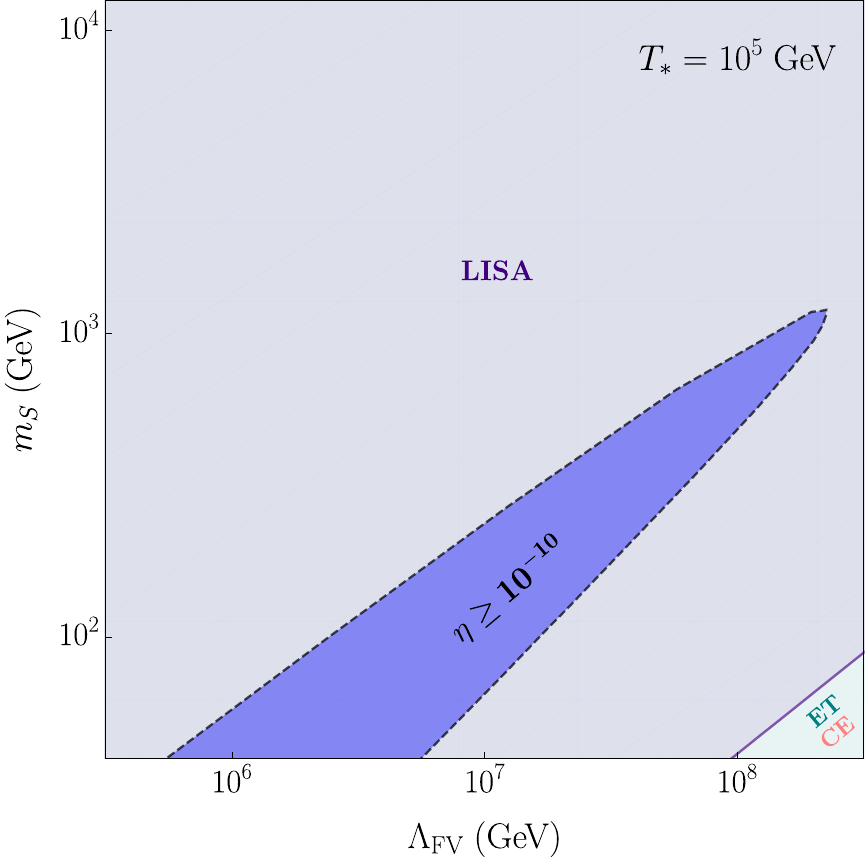}
    \caption{\it }
    \end{subfigure}\hfill
    
\caption{ \small{Same as Figure \ref{fig:SNR8} but for $n_T=0.5$. }}
    \label{fig:SNR5}
\end{figure}


In Figure~\ref{fig:SNRplot} we show the SNR calculation expected from primordial GWs with an era of flavon domination for U-DECIGO, BBO and $\mu$-ARES detectors as representative examples. (The signal power spectrum is discussed in the previous section in detail.) The signal increases with increasing flavon mass and decreasing $\LFV$, since the power spectrum is less suppressed as can be seen in Equation~\ref{eq:Rsup}. Nevertheless, even with a suppressed spectrum, a significant part of the parameter space can be covered by these proposed detectors.

In Figures  \ref{fig:SNR8} and \ref{fig:SNR5} we highlight the parameter space where successful flavon baryogenesis is possible and show the future sensitivities of different GW experiments in the parameter space $10~{\rm GeV} \lesssim m_S \lesssim 10$ TeV and $10^5~{\rm GeV} \lesssim \LFV \lesssim 10^{9}$ GeV. We find that GW detectors will not be sensitive to this parameter region for $n_T=0$ if the flavon domination temperature is high, $T_\ast=10^8$~GeV. When the flavon domination temperature is smaller, e.g.  $T_\ast=10^5$~GeV, the power spectrum is less suppressed and U-DECIGO will be able to cover part of the parameter space for successful flavon baryogenesis even for $n_T=0$. In the case of a blue-tilted spectrum, e.g. $n_T=0.5$, many experiments including LISA will be sensitive to the parameter region with successful baryogenesis. 

\subsection{Compatibility with NANOGrav Results}\label{sec:NG}

So far we have taken the tensor-to-scalar ratio $r=0.035$, the maximum allowed value from CMB measurements~\cite{BICEP:2021xfz}, to maximize the amplitude of the GW spectrum for the $n_T$ values we considered. These measurements allow for smaller $r$. At the same time, it has been shown that a primordial GW spectrum can explain the recent NANOGrav data~\cite{NANOGrav:2023gor} for the best-fit parameters of $r\sim10^{-14.06\pm5.82}$, $n_T=2.61\pm0.85$, $T_{\rm RH}\sim1-10$ GeV~\cite{NANOGrav:2023hvm}. The minimal flavon baryogenesis scenario we studied here does not work if $T_{\rm RH}\lesssim 160 $~GeV since the sphalerons would never have been active in the early universe. In this case, one needs to explore different mechanisms to produce the baryon asymmetry.

\section{Conclusion}
\label{sec:conclusion}

In this work we analyzed the sensitivity of several GW detectors to the flavor violation scale and an associated baryogenesis scenario. The flavon field, which is introduced in many new physics models to explain the hierarchy of fermion masses in the SM, is expected to dominate the energy density of the universe leading to a period of matter domination before decaying away. During this era of (intermediate) matter domination driven by the flavon field, right-handed electrons stay out thermal equilibrium until sphalerons turn off at $T_{\rm EW}\simeq 160$~GeV. Therefore, an asymmetry in the electrons created by flavon decays can survive and be transferred to a baryon asymmetry and may explain the observed matter-antimatter asymmetry of the universe. 

The period of flavon-domination and the subsequent entropy injection due its decay also lead to spectral features in the primordial GWs produced by metric tensor fluctuations generated during inflation. Specifically, we showed that primordial GW amplitude is suppressed above a frequency $f_{\rm sup}$ determined by the temperature of flavon decay, which in turn is governed by the flavon mass $m_S$ and the effective scale of flavor violation $\LFV$. This suppression can be seen in Figure~\ref{fig:spectrum} for several model parameters. For typical flavon decays at $T\sim $ GeV, the power spectrum is suppressed above $f_{\rm sup} \sim 10^{-8}$ GeV. If the flavon lifetime is much shorter, deviation from the unsuppressed standard GW spectrum is less important. The amount of suppression can be quite significant, \emph{e.g.} $O(10^8)$ for the same parameters. (See equations \ref{eq:freq} and \ref{eq:Rsup}.) Even with this suppression, the resulting primordial GW spectrum along with spectral features is detectable in various future GW missions as we have extensively shown. 

We calculated the signal-to-noise ratio for future detectors like U-DECIGO, BBO, LISA and $\mu-$ARES for 4 years of observation time. Taking SNR$\geq 10$ as a detection benchmark, we identified the parameter space for $m_S$ and $\LFV$ that both gives the correct baryon asymmetry and that can be detected in various GW experiments, as shown in Figures \ref{fig:SNR8} and \ref{fig:SNR5}. This detectable baryogenesis region corresponds to $10~{\rm GeV}\lesssim m_S \lesssim 10~$TeV and $10^5~{\rm GeV}\lesssim\LFV\lesssim 10^9$~GeV for a flavon domination temperature of $T_\ast = 10^5$~GeV and a spectral index $n_T\geq 0$. For larger $T_\ast$, e.g. $T_\ast = 10^8$~GeV, the GW spectrum is suppressed more and hence it is harder to detect in many experiments. In order to probe the region with successful baryogenesis, a blue-tilted spectrum is needed. Indeed, we note that in most of the parameter space, for detectability,  $n_T>0$ is needed. This is a deviation from what is expected in minimal single-field, slow-roll inflation which predicts $n_T \approx -r/8$ as prevalent in several cosmological scenarios \cite{Brandenberger:2006xi,Baldi:2005gk,Kobayashi:2010cm,Calcagni:2004as,Calcagni:2013lya,Cook:2011hg,Mukohyama:2014gba,Kuroyanagi:2020sfw}.

Gravitational wave astronomy aspires to achieve precisions that are orders of magnitude better than the current detectors. This new era of GW detectors worldwide will make the dream of testing fundamental BSM mechanisms, e.g. for flavor physics, matter-antimatter asymmetry of the universe and and inflationary cosmology, a reality in near future.


\section*{Acknowledgements}
ZAB is supported by DST-INSPIRE fellowship. SI is supported by the Natural Sciences and Engineering Research Council (NSERC) of Canada. SI and AG are grateful to the Mainz Institute for Theoretical Physics (MITP) of the Cluster of Excellence PRISMA+ (Project ID 390831469) for hosting the New Proposals For Baryogenesis workshop, during which this work was started.

\bibliographystyle{JHEP}
\bibliography{ref}
\end{document}